\documentclass{elsarta}
\usepackage{graphicx}
\begin{document}

\hspace*{4 in}CUQM-121\\
\vspace*{0.4 in}
\begin{frontmatter}
\title{Solutions to the 1d Klein-Gordon equation with  cutoff Coulomb potentials}
\author[hall]{Richard L. Hall}
\address[hall]{Department of Mathematics and Statistics, Concordia University,
1455 de Maisonneuve Boulevard West, Montr\'eal,
Qu\'ebec, Canada H3G 1M8}
\begin{abstract}
In a recent paper by Barton (J. Phys. A {\bf 40}, 1011 (2007)), the 1-dimensional Klein-Gordon equation was solved analytically for the non-singular Coulomb-like potential $V_1(|x|) = -\alpha/(|x|+a).$  In the present paper, these results are completely confirmed by a numerical formulation that also allows a solution for an alternative cutoff Coulomb potential $V_2(|x|) = -\alpha/|x|,$ ~$|x| > a,$ and otherwise $V_2(|x|) = -\alpha/a.$  
\end{abstract}

\end{frontmatter}
\section{Introduction}\label{intro}
We consider the Klein-Gordon equation in one dimension, with an attractive vector potential $V(|x|)$ (the time component of a 4-vector), and mass $m$.  In units in which $\hbar = c = 1,$ the equation may be written~\cite{greiner} in 
the form
\begin{equation}\label{eq1}
(-D^2+m^2)\psi(x)=(E-V)^2\psi(x),
\end{equation}
where $E$ denotes the energy, and, for bound states, $\psi\in L^2(R).$  Barton~\cite{barton} has recently presented a detailed study of (\ref{eq1}) by analytical methods in the case that $V$ is a non-singular Coulomb-like potential with the form
\begin{equation}\label{eq2}
V_1(|x|) = -\alpha/(|x|+a),
\end{equation}
where $\alpha$ is the fine-structure constant, and $a$ is a positive cutoff parameter (called $R$ by Barton). We shall adopt the approximation for $\alpha$ and the definition of the parameter $\delta$ given by
$$\alpha = \frac{1}{137}\quad{\rm and}\quad \delta := 1/2 - \sqrt{1/4 -\alpha^2}\approx5.32821 \times 10^{-5}.$$
Barton's paper augments the classic work by Loudon~\cite{loudon} on the corresponding nonrelativistic problem.  There is an extensive literature related to this physical system which is reviewed and discussed in Barton's paper. An interesting discussion was presented in an earlier paper by Dombey~\cite{dombey}, who studied so-called hydrino states and the Klein-Gordon and Dirac Coulomb problems in three and two dimensions respectively. 

Since the potential $V(|x|)$ is symmetric, the eigenfunctions of (\ref{eq1}) can be taken to be even or odd. Most of the energies are associated with a set of odd-even doublets, which are essentially Balmer like and may be expressed~\cite{barton} for small $a$ by the formula
\begin{equation}\label{eq3}
\frac{E_n}{m} \approx \left\{1 + \left(\frac{\alpha}{n+\varepsilon-\delta}\right)^2\right\}^{-\half} = 1 - \frac{\alpha^2}{2(n+\varepsilon-\delta)^2} + {\mathcal O}(\alpha^4),
\end{equation}
where  $n = 1,2,3,\dots,$ and small-$a$ approximations for $\varepsilon$ are given by
$$\varepsilon({\rm odd})\approx 2\alpha m a\quad{\rm and}\quad \varepsilon({\rm even}) \approx \frac{2}{\left[\alpha/ma + 2\log(n/2\alpha m a)\right]}.$$  
The eigenvalues expressed by these analytical approximations of Barton are are not difficult to verify numerically.  However, the principal goal of the present paper is to determine, not the Balmer-like energies, but the bottom of the spectrum of (\ref{eq1}), that is to say, the energy of the `anomalous' even state. Guided by the careful mathematical analysis of Ref.~\cite{barton} which has exhibited the key spectral features generated by the potential (\ref{eq2}), we have been able to devise a purely numerical search procedure which is then not limited to any particular design for the cutoff potential.  We shall discuss this matter more fully in the next section.  First we define the spectral parameters used by Barton: we adopt these for the present paper so that our results will be immediately comparable. Thus we define:
\begin{equation}\label{eq4}
s := \frac{m a}{\delta} \quad {\rm and}\quad\beta := \frac{\alpha E}{\delta\sqrt{m^2-E^2}}.
\end{equation} 
The idea is this: the coupling $\alpha$ is fixed; we choose $m$ and the cutoff radius $a$, and this determines $s$; the Klein-Gordon equation (\ref{eq1}) is now completely specified and is solved for $E$ for the chosen state; finally $\beta(s)$ is found from $E$ and Eq.(\ref{eq4}).  For any other potential, also having a single cutoff parameter $a$ (and obeying the scaling law discussed below), we may continue to use the same definitions. We observe that
$$\frac{\partial s}{\partial a} = \frac{m}{\delta} > 0 \quad {\rm and}\quad \frac{\partial \beta}{\partial E} = \frac{\alpha m^2}{\delta (m^2-E^2)^{\frac{3}{2}}} > 0.$$
These monotonicities mean that we can safely think of the pair $\{s,\ \beta\}$ as scaled images of the pair $\{a,\ E\}.$  This observation is important since one of the interesting and inherent features of the problem we study is that the parametric curve $\{s,~ \beta(s)\}$, $s > 0,$  has two branches; we have to be sure {\it a priori} that this property is not merely a consequence of the definitions of the parameters.

Another common feature exhibited by the Klein-Gordon equation, with either of the Coulomb-like potentials discussed in this paper, is to do with scaling.  If we replace $x$ by $x' = x/\sigma$, $\sigma > 0,$ re-write the equations, and then choose the scale $\sigma = 1/m$, we can easily show that, for each fixed coupling $\alpha >0,$ the energy $E$ depends on the remaining parameters $\{a,\ m\}$ according to the scaling law
$$\frac{E}{m} = F(m a),$$
in which $F$ may have more than one branch. A consequence of this law is that it is sufficient for the determination of the spectral function $F$ to consider the problem with $m$ fixed, say $m = 1.$  This reduction in complexity is particularly important for a numerical approach. Thus, since $\alpha $ is in any case constant, what we need to do essentially is to determine how $E$ depends on $a,$ or, equivalently, how $\beta$ depends on $s.$
\section{The $\beta(s)$ curves}\label{beta}
We use fundamental numerical methods to study the lowest even eigenstate of (\ref{eq1}).  We make an initial  guess for $E$ in the range $-m = E_L < E < E_U = m$ and (for the even solution) set $\psi(0) = 1$ and $\psi'(0) = 0;$ for discrete eigenvalues, the wave function must be normalizeable and, in particular,  satisfy $\lim_{x\rightarrow \infty}\psi(x) = 0$ (although $\psi$  need not be normalized).  We then integrate numerically (by using a Runge-Kutta method with local error $\sim h^5$) from $x=0$, with increasing $x$.  We follow the size of $\psi(x)$ and count the nodes. 
Experience with Schr\"odinger and Sturm-Liouville eigenproblems indicates that, if $E$ is too small, then $|\psi(x)|$ becomes large without exhibiting the requisite number of nodes; meanwhile, if $E$ is too large, there are too many nodes. Thus the energy bounds $\{E_L,~ E_U\}$ can be repeatedly and more narrowly re-chosen on the basis of the observed numerical behaviour of $\psi(x)$ as $x$ increases.  This simple logic works for the Klein-Gordon equation too, provided the energy sought is positive; when the energy sought is negative, too many nodes indicates instead that $E$ is too small.  This experimental observation allows us to find the lower branch of the $\beta(s)$ curve. To be still more explicit on this point, for the upper branch we search in the patch $(0,\ m),$ and, for the lower branch in the patch $(-m,\ 0),$ and we use the appropriate relation between the node count and the size of $E$ mentioned above.  We note in passing that the interesting details of the graph for points near to $\beta = 0$ necessistated a distinct and possibly superior approach that is described below. Our graphical results for the potential $V_1$ studied by Barton are shown in Figure~1.
\begin{figure}[htbp]
\centering
\includegraphics[width=12 cm]{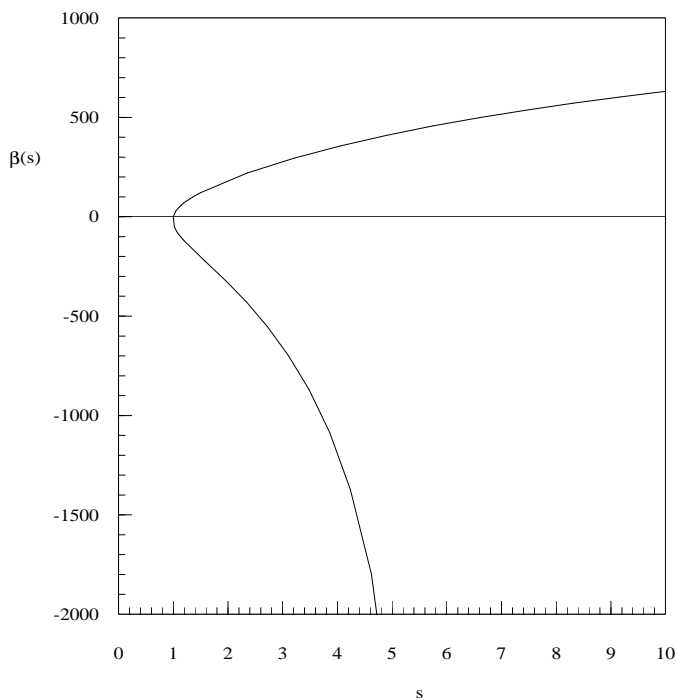}
\caption{Plot of $\beta(s)$ for the potential $V_1(|x|) = -\alpha/(|x|+a).$}
\label{Fig. (1)}
\end{figure}
We confirm the special values $s_0$ and $s_{\infty}$ (corresponding to Barton's $s_2$ and $s_4$) satisfying 
$$\beta(s_0) = 0\quad {\rm and} \quad \lim_{s\rightarrow s_\infty} \big( \beta(s)\big) = -\infty.$$
In terms of $E$ and $a$, this limit corresponds to the supercritical limit $E\rightarrow -m.$ We find for the potential $V_1$ that $s_0 = 0.99906868$ and $s_{\infty} = 6.1711.$ 
\clearpage
 Figure~1 shows the overall shape of the $\beta(s)$ curve but, as Barton found, a magnified patch near $\beta = 0$ is needed to complete the presentation, and indeed to determine the minimum value $s_m = 0.99136$ of $s$: we show this enlarged view in Figure~2. We note that Barton writes $s_m = s_3$, and also that $s= s_m$ corresponds to $a = a_m,$ where $ma_m = 5.28217\times 10^{-5}.$   The earlier search policy based on the sign of $E$ and described above was perfectly satisfactory for the large-scale features of the graph shown in Figure~1.  However, in order to generate Figure~2, we were obliged to accepted the fact that $a$ is a function of $E$ (and not {\it vice versa}), even when $E$ does not change sign. Thus, for each point on the graph, we fixed $E$ in the patch $(-m,\ m)$ and searched for the corresponding value of $a$ that would yield a wave function $\psi\in L^2.$  We found, for either sign of $E,$ that the number of nodes increased with {\it decreasing} $a$; we are only interested, of course, in zero nodes, but the computer program must `know' how to adjust $a$ on the basis of the values of $\psi(x)$ for large $x.$  At the end we converted to the $\{s,\ \beta\}$ pair.  This approach, which is independent of the sign of $E$, appears to be robust and indeed might well be used for the all the points on the graph. 
\begin{figure}[htbp]
\centering
\includegraphics[width=12 cm]{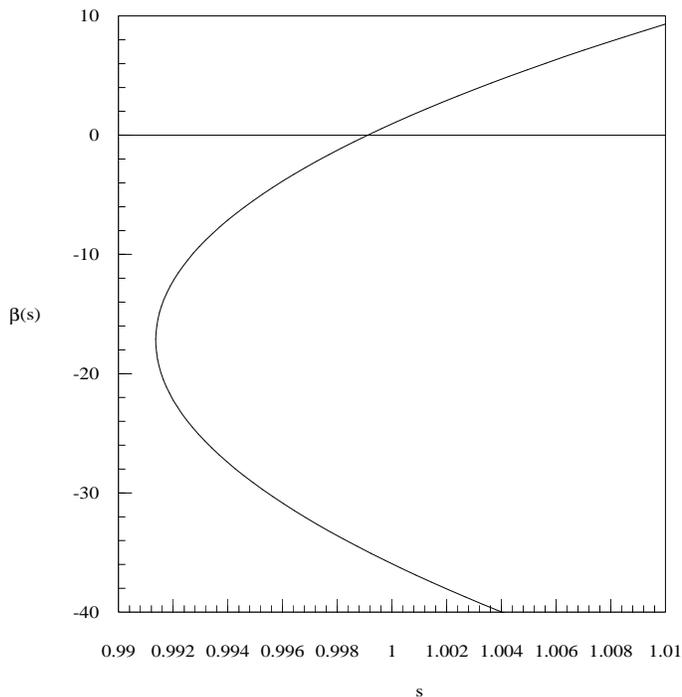}
\caption{Enlarged view of $\beta(s)$ for the potential $V_1(|x|) = -\alpha/(|x|+a).$ This graph shows the minimum value $s_m = 0.99136$.}
\label{Fig. (2)}
\end{figure}
The shape of the graph of $\beta(s)$ and the special values $\{s_0,\ s_{\infty},\ s_m\}$ completely confirm the analytical approximations of Barton~\cite{barton} for this problem.

We now consider an alternative $V_2$ to the cutoff potential given by
\begin{eqnarray}\label{eq5}
V_2(|x|)=\left\{
\begin{array}{ll}
-\frac{\alpha}{|x|},\quad & |x| > a,\\
-\frac{\alpha}{a},\quad & |x| \leq a.\\
\end{array}
\right.
\end{eqnarray}
The corresponding graphs of $\beta(s)$ obtained by solving (\ref{eq1}) with this potential are shown in Figs.~(3)~and~(4).
\begin{figure}[htbp]
\centering
\includegraphics[width=12 cm]{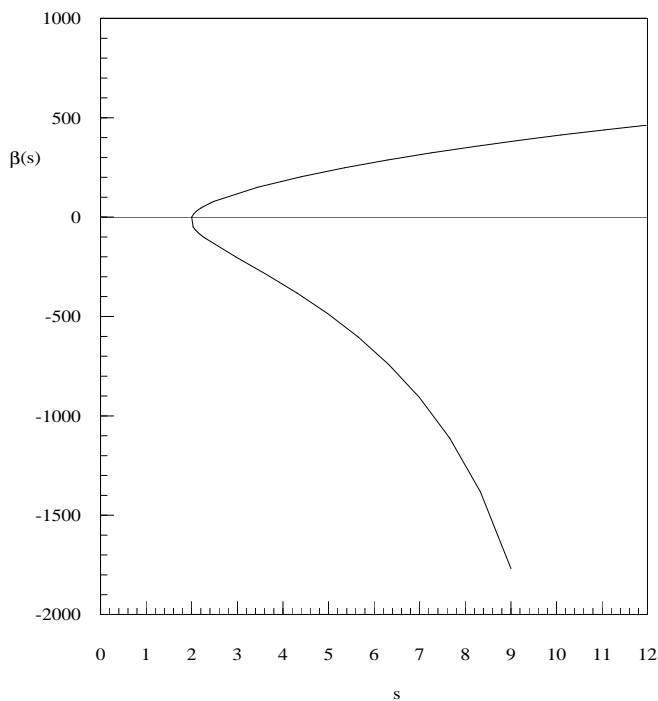}
\caption{Plot of $\beta(s)$ for the cut-off Coulomb potential $V_2(|x|) = -\alpha/|x|,$~ $ |x| > a,$ 
otherwise $V_2(|x|) = -\alpha/a.$}
\label{Fig. (3)}
\end{figure}
\begin{figure}[htbp]
\centering
\includegraphics[width=12 cm]{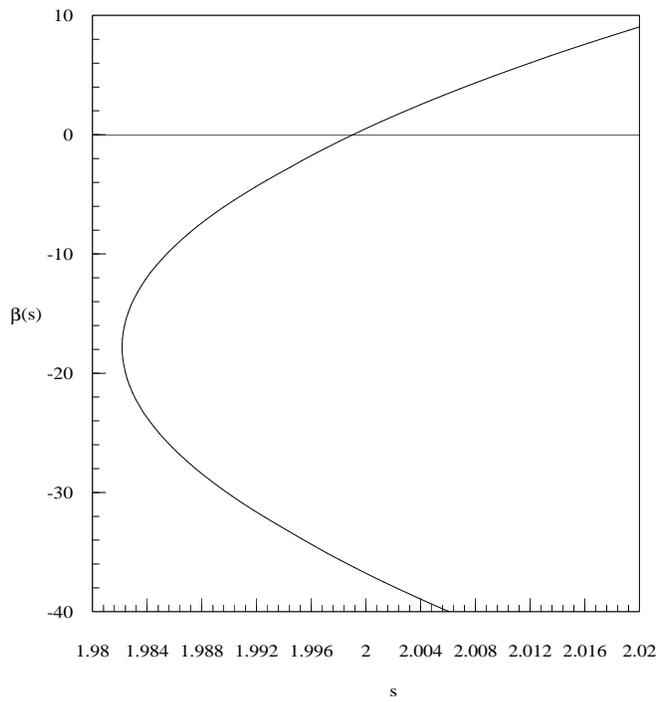}
\caption{Enlarged view of $\beta(s)$ for the cut-off Coulomb potential $V_2(|x|) = -\alpha/|x|,$~ $ |x| > a,$ 
otherwise $V_2(|x|) = -\alpha/a.$  This graph shows the minimum value $s_m = 1.98216$.}
\label{Fig. (4)}
\end{figure}

The corresponding special values are $s_0 = 1.9982289$, $s_{\infty} = 11.9777,$ and, from Figure~4, the minimum allowed value $s_m = 1.98216$ (corresponding to $m a_m = 1.05614\times 10^{-4}$). 

\clearpage

\section{Conclusion}\label{conc}
It should be emphasized that all the results in this paper have been obtained directly from Eq.(\ref{eq1}) and are therefore completly independent of the formulation chosen by Barton for the mathematical analysis appropriate to $V_1;$   the confirmation is all the more significant because of this independence.  Since we obtain similar results for an alternative cutoff potential $V_2$, the main features of the solution, discussed at length by Barton,  are likely to be independent of the particular form of the cutoff.  This paper illustrates a constructive interplay between the analytic and numerical work: the complicated two-branch relationship between the energy $E$ and the cutoff radius $a$ was straightforward to follow numerically once it's general form had been found analytically for one example.  Without further mathematical work it is reasonably safe to assume that the key features represented by the curves $\beta(s)$ are generic to one-parameter cutoff Coulomb potentials satisfying the scaling rule $E/m = F(ma):$ in particular, no discrete spectrum would be expected in the potential limit $a\rightarrow 0.$   

\section{Acknowledgements}
\medskip
\noindent Partial financial support of this work under Grant No. GP3438 from the 
Natural Sciences and Engineering Research Council of Canada
 and hospitality of the Department of Mathematics of the University of Auckland, 
where some of this work was carried out, is gratefully acknowledged.


\begin{thebibliography}{00}
\bibitem{greiner}W. Greiner, {\it Relativistic Quantum Mechanics: Wave Equations\/}, 3rd ed. (Springer, Berlin 2000).
\bibitem{barton}
G. Barton, J. Phys. A: Math. Gen. {\bf 40}, {1011 (2007)}.
\bibitem{loudon}
R. Loudon, Am. J. Phys. {\bf 27}, {649 (1959)}.
\bibitem{dombey}
N. Dombey, Phys.  Lett. A {\bf 360}, {62 (2006)}.


\end{thebibliography}
\end{document}